\documentclass[pra,twocolumn,showkeys,showpacs,10pt,superscriptaddress,aps]{revtex4-1}
\usepackage[english]{babel}
\usepackage{amsmath}
\usepackage{dcolumn}
\usepackage[english]{babel}
\usepackage{amssymb}
\usepackage{dsfont}
\usepackage{textcomp}
\usepackage{graphicx,color}
\usepackage[utf8]{inputenc}
\usepackage[normal]{subfigure}
\usepackage{lipsum}
\usepackage{caption}

\usepackage{mathptmx}

\newcommand{\ie}{i.\thinspace{}e., }
\newcommand{\be}{\begin{equation}}
\newcommand{\ee}{\end{equation}}
\newcommand{\bea}{\begin{eqnarray}}
\newcommand{\eea}{\end{eqnarray}}

\renewcommand{\vec}{\mathbf}

\newcommand{\ave}[1]{\langle #1 \rangle}
\newcommand{\avee}[1]{\langle\!\langle #1 
\rangle\!\rangle}

\newcommand{\comut}[2]{[#1,\,#2\,]}

\usepackage{units}
\graphicspath{{Skizzen/}}

\begin{document}
 \title{Tailored quantum statistics from broadband states of light }

\newcommand{\affT}{Institute of Applied Physics, Technical University Darmstadt, Germany}
\newcommand{\affC}{also with Center of Smart Interfaces, Technical University Darmstadt, Germany} 

\author{S. Hartmann} 
\email{sebastien.hartmann@physik.tu-darmstadt.de}
\affiliation{\affT}

\author{F. Friedrich}
\affiliation{\affT}

\author{A. Molitor}
\affiliation{\affT}

\author{M. Reichert}
\affiliation{\affT}

\author{W. Els\"aßer}
\affiliation{\affT}
\affiliation{\affC}

\author{R. Walser}
\affiliation{\affT}

\date{\today}

\begin{abstract}
We analyze the statistics of photons originating from amplified spontaneous emission generated by a quantum dot superluminescent diode. Experimentally detectable emission properties are taken into account by parametrizing the corresponding quantum state as a multi-mode phase-randomized Gaussian density operator.
The validity of this model is proven in two subsequent experiments using fast two-photon-absorption detection observing second order equal-time- as well as second order fully time-resolved intensity correlations on femtosecond timescales. In the first experiment, we study the photon statistics when the number of contributing longitudinal modes is systematically
reduced by applying well-controlled optical feedback. In a second experiment, we add coherent light from a single-mode laserdiode to  quantum dot superluminescent diode broadband radiation. Tuning the power ratio, we realize tailored second order correlations ranging from Gaussian to Poissonian statistics. Both experiments are very well matched by theory, thus giving first insights into quantum properties of radiation from quantum dot superluminescent diodes.
\end{abstract}


\maketitle

\section{Introduction}
The intriguing mechanism of amplified spontaneous emission (ASE) 
results in broad radiation spectra, 
high output intensities 
as well as strong directionality of emission \cite{LeeBurrusMiller1973,Siegman1986,Wiersma2008}. 
Shortly after the invention of the laser in 1960, 
ASE has been subject of intense theoretical and experimental coherence studies, 
in particular due to the disturbing influence of ASE at gas laser threshold 
\cite{Prescott1964,Arecchi1965,Arecchi1966,Smith1966}. 
Later in the 1970s, Allen and Peters were the first to address the 
ASE phenomenon, putting it into context to Dicke's superradiance \cite{Allen1970}.
They defined ASE as “highly directional radiation emitted by an extended medium 
with a randomly prepared population inversion in the absence of a laser cavity”.
Supported by theoretical studies and He-Ne-gas discharged tube amplification experiments, 
they established the ASE threshold condition, the pump-output-intensity behavior, saturation effects and spatial coherence properties \cite{Peters1971a,Peters1971b,Peters1971c,Peters1972d}.

Nowadays, well-developed and highly sophisticated semiconductor laser technology
provides compact ASE light sources realized with superluminescent diodes (SLD). 
They are semiconductor-based opto-electronic emitters generating broadband light.
The technological development of these high performance devices 
with wide-ranging material structure systems is boosting application areas such as telecommunication, medical and industrial application \cite{Huang1991,Urquhart2007,Judson2009,Velez2005}. Especially when it comes to the need of compact, miniaturized light sources with spectrally broad properties, SLDs are a first choice.
To foster the technological progress, it is indispensable
to develop theoretical models of SLD emission in close 
adaption to specific material systems targeting specific device properties such as
pulse performances \cite{Majer2011}, 
amplification improvements \cite{Gioannini2011} and 
noise behavior \cite{McCoy2005,Marazzi2014,Liu2011}. Sophisticated 
numerical models based on rate equations and 
travelling wave approaches are developed \cite{Uskov2004,Gioannini2013,Rossetti2011}
and guide future progress.

However, fundamental quantum optical studies on SLD light emission, particularly regarding higher order coherence properties have not - to the best of our knowledge - been addressed so far. 
The complex material structures with predominantly application driven objectives often lead to theoretical approaches ignoring the quantum aspect of the light state. 
In this context, photon statistics is the footprint of the quantum nature
of light, directly related to the emission process and quantified by the 
central degree of second order coherence 
$g^{(2)}(\tau=0)$ \cite{MandelWolf,Degiorgio2013}.
It is important to point out that experimental access to photon statistics via the determination of the second order intensity auto-correlation function $g^{(2)}(\tau)$ in this spectrally  ultrabroadband regimes, was not possible until 2009.
Then, Boitier and coworkers \cite{Boitier2009} demonstrated via two-photon-absorption (TPA) in a semiconductor-based photocathode of a photomultiplier, evidence of the photon bunching effect on the corresponding ultrashort timescales.
Since then, the second order correlations $g^{(2)}(\tau)$ of broadband light states can be globally resolved, in the sense that all contributing spectral components are simultaneously detected. 
A number of investigations have exploited this elegant TPA detection technique so far, 
referring to characterizations and applications of broadband semiconductor emitters 
regarding their photon statistical characteristics \cite{Hartmann2013,Blazek2012,Blazek2011a,Jechow2013,Boitier2011,Boitier2013,Nevet2013}.

Moreover, the exploitation of quantum dot (QD) based gain material in SLD structures, enables on the one hand, a strong enhancement of the spectral broadening \cite{StranskiKrastanov1938} and introduces on the other hand, a non-negligible quantum aspect for the carrier dynamics in the semiconductor material as well as for the generation of photons \cite{Sun1999}. 
The quantized zero-dimensional carrier systems of the inhomogeneously broadened quantum dots in SLD structures generate a strong emission state hierarchy \cite{Stier1999}, which has only recently been extensively investigated regarding its impact on the coherence properties \cite{Blazek2012}. Recent studies on QD-SLD light coherence \cite{Blazek2011a} have revealed a temperature induced reduction of the intensity correlations while the ultrabroadband spectral emission maintains unchanged.
This novel hybrid light state exhibits very low first order coherence as it is 
spectrally broad in term of wavelength $\Delta\lambda=\unit[65]{nm}$ or angular frequency 
$b=\unit[2\pi\cdot12.41]{THz}$, but shows suppressed $g^{(2)}(0)=1.33$ laser-like intensity correlations. 

These latest experiments require the development of a quantum theory of ASE light states emitted by QD-SLDs. In this contribution, we propose a simple model in 
Sec.~\ref{sec2}, which allows to include specific emission properties of a given QD-SLD device without considering specific structural characteristics. In particular, we surmise a multi-mode phase-randomized Gaussian (PRAG) quantum state and discuss the evaluation of moments as well as correlation functions of the light field.
In order to probe this hypothesis, we match it with observations in two different types of experiments. Results of the first experiment are reported in Sec.~\ref{opticalfeedback}, where the number of modes of the  QD-SLD light is varied systematically via optical feedback and we observe the response in the photon statistics.
A second experiment is presented in Sec.~\ref{heterodyne}, where we induce a transition in the photon statistics by superimposing coherent light from a laserdiode with
broadband emission of a QD-SLD. 
Our conclusions and future perspectives are presented in Sec.~\ref{conc}. We postpone technical aspects of interpolating spectra as well as the Euler-Maclaurin formula to two appendices.

\section{Emission from a quantum-dot SLD}
\label{sec2}
The emission of an edge-emitting quantum-dot superluminescent diode is described by the transversal electric field 
$\hat{{\vec E}}=\hat{\vec E}^{(+)}+\hat{\vec E}^{(-)}$. In order to model a broad radiation spectrum, we need to consider a superposition of numerous longitudinal modes $N$ for the positive frequency part of the electric field 
 \begin{align}
  \label{1.1}
  \hat{\vec E}^{(+)}(t,\vec r)&=\hat{\vec E}^{(+)}(x,y,z-ct)=\sum_{j=1}^{N} \mathcal{E}_j u_j(t,\vec r) \hat{a}_j
 \end{align}
at position $\vec r=(x,y,z)$ and time $t$. 
The structural composition of QD-SLDs \cite{zhang2007effect} enforces a linear $y$-polarization upon the field.
As we are interested in the forward propagating field, we want to consider the spatio-temporal modes of the field
$u_j=\chi(x,y)e^{i(k_j z-\omega_j t)}\vec e_y$. They are formed by a single transverse wave function $\chi$, as well as longitudinal plane waves with  wave numbers $k_j=2\pi j/L$. Here, $L$ is the length of the optical system and $A$ is the cross section area. Then, the  mode functions are normalized to the volume $V=AL$, 
\begin{align}
\label{normalization}
  \int_V \text{d}^3r\,|u_j(t,\vec{r})|^2&=V.
\end{align}
The quantized amplitude $\hat{a}_j$ of the electromagnetic field annihilates photons of mode $j$ and satisfies the bosonic commutation relation $\comut{\hat{a}^{ }_i}{\hat{a}^\dagger_j}=\delta_{ij}$. This field is an approximate solution of the free Maxwell equation with a linear dispersion relation $\omega_j=c k_j$ with the velocity of light $c$. The field normalization $\mathcal{E}_j=i\sqrt{\hbar\omega_j/2\epsilon_0 V}$ of Eq.~\eqref{1.1} is chosen such that the energy of the transversal field is given by
\begin{align}
  \label{hamiltonian}
  \hat H=\sum_{j=1}^{N} \hbar\omega_j^{ }\hat a_j^\dagger\hat a_j^{ },
\end{align}
where $\epsilon_0$ is the vacuum permittivity.  
\subsection{Quantum state of the electro-magnetic field}

\begin{figure}[h]
    \includegraphics[width=\columnwidth]{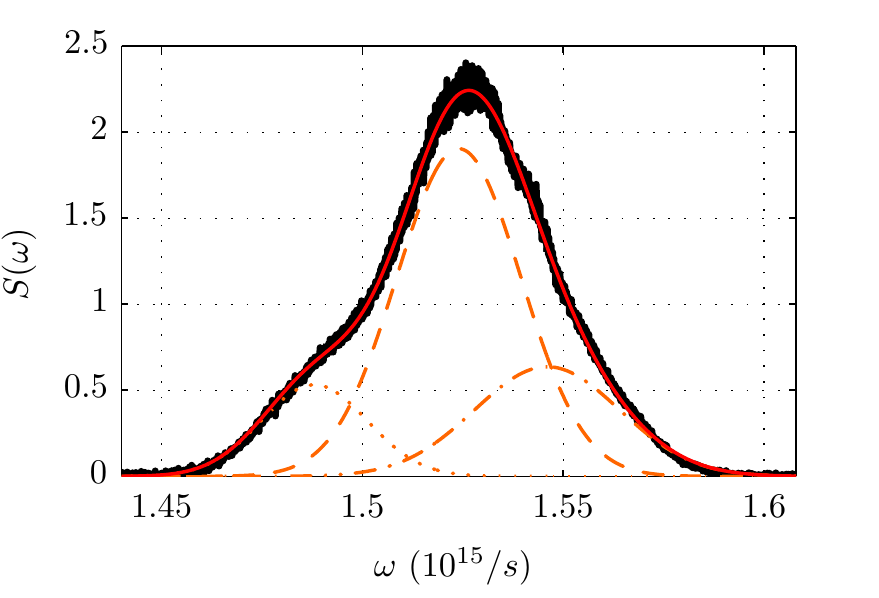}
    \caption{Measured optical power spectrum $S(\omega)$ in arbitrary units versus angular frequency $\omega$ of the QD-SLD with central frequency $\bar\omega=\unit[2\pi\cdot242.6]{THz}$ and spectral width $b=\unit[2\pi\cdot13]{THz}$. A three term Gaussian fit (red, solid line) exhibits a dominant central emission line (orange, dashed line), as well as a lower side band (orange, dotted line) and an upper side band (orange,dashed dotted line).}
    \label{spectrum}
\end{figure}

In order to parametrize the quantum state of the QD-SLD emission, we consider an observed optical spectrum $S(\omega)$, shown in Fig.~\ref{spectrum}, as an input. Clearly, the diode emits on a central angular frequency 
$\bar\omega=\unit[2\pi\cdot242.6]{THz}$ ($\bar\lambda=\unit[1236]{nm}$) showing a Gaussian shaped distribution with a very broad spectral width of $b=\unit[2\pi\cdot13]{THz}$ . 
There are also upper and lower side-bands visible, whose strength can be quantified by a three term Gaussian interpolation of the data (cf. Tab.~\ref{tablefit} of appendix \ref{fita}) \cite{Grundmann2002}.

This obviously demonstrates that the quantum state cannot be described by thermal Planck distribution
and that the broadband emission is strongly 
incoherent as measured by the first order correlation function $g^{(1)}(\tau)$. 
Regarding the intensity correlations, QD-SLD emission can exhibit significant deviations from an ideal thermal photon statistics $g^{(2)}(0)=2$. 
A reduction down to laser-like values of  $g^{(2)}(0)=1.33$ at temperatures around 
$T=\unit[190]{K}$
has been measured \cite{Blazek2011a}. This can be interpreted as a delicate balance between spontaneous and stimulated emission in QD-SLDs. 

These experimental facts about the amplified spontaneous emission of the device are captured by the multi-mode phase-randomized Gaussian (PRAG) state  \cite{mollow68,allevi2013,Bondani2009,Bondani2009Second}
\begin{eqnarray}
   \label{statesld}
   \hat \rho_{s}=\frac{1}{(2\pi)^N}\int_{0}^{2\pi}
	\text{d}^N\phi \,\hat D(\gamma)\hat \rho_{T}\hat D^\dagger(\gamma)
  \end{eqnarray}
with the multi-mode displacement operator
\begin{eqnarray}
 \label{1.4}
  \hat D(\gamma)=e^{\sum_{i=1}^{N}(\gamma_i^{ }\hat a_i^\dagger-\gamma_i^\ast \hat a_i^{ })}.
\end{eqnarray}
A natural choice for an equilibrium state $\hat\rho_T$ is the canonical operator
\begin{eqnarray}
 \label{thermal}
 \hat \rho_{T}=\frac{e^{-\beta\hat H}}{Z},
\end{eqnarray}   
where $Z=\text{Tr}\{e^{-\beta\hat H}\}$ is the canonical partition function and $\beta=1/k_B T$ is proportional to the inverse temperature. 

\begin{figure}[h!]
    \includegraphics[width=.3\textwidth]{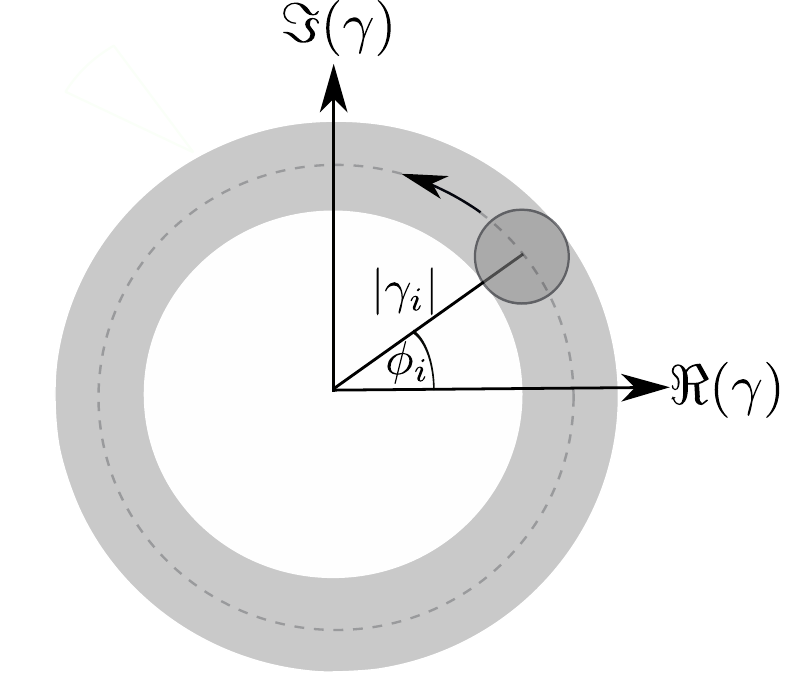}
    \caption{Schematic phase-space representation of a phase-randomized Gaussian state $\hat \rho_{s}$. We depict the mode 
		$i$, which is prepared in a thermal state, displaced by $\gamma_i=|\gamma_i|e^{i\phi_i}$ 
		and all phase angles are randomized according to Eq.~\eqref{statesld}.}
    \label{states}
\end{figure}

A phase-space representation of this 
PRAG state is shown 
in Fig.~\ref{states}. There, we consider a generic mode $i$. Starting from a Gaussian state centered at the origin, we shift it by a complex amplitude $\gamma_i=|\gamma_i|e^{i\phi_i}$ and randomize the phases $\phi_i$, finally.

It is instructive  to consider the limit of vanishing temperature $T\rightarrow 0$. There, one finds for the probability of finding $n$ photons in mode $i$, 
\begin{align}
  \label{poisson}
  p_{i}(n)=\ave{\delta(n-\hat{n}_i)}=
	e^{-|\gamma_i|^2}\frac{|\gamma_i|^{2n}}{n!}.
\end{align}
As usual, quantum averages 
$\langle\ldots\rangle=
\text{Tr}\{\ldots\hat{\rho}_s\}$ requires tracing over the state.
Clearly, this coincides with the Poissonian distribution of a coherent state 
$|\gamma\rangle=\hat D(\gamma)|0\rangle$, even though we have completely randomized the phases of this incoherent state of Eq.~\eqref{statesld}. 
\subsection{Ensemble averages}
The properties of the PRAG state are completely characterized by first and second order moments
\begin{align}
  \label{ave}
  \ave{\hat{a}_l}=0,\qquad
  \ave{\hat{a}_j^\dagger\hat{a}_i}=\left(|\gamma_i|^2+n_T(\omega_i)\right)\delta_{ij}.
\end{align}
All higher order moments can be determined by Wick's theorem. Here, the mean thermal occupation number
\begin{eqnarray}
  \label{1.7}
  n_{T}(\omega)=\frac{1}{e^{\beta\hbar\omega}-1}
\end{eqnarray}
is given by the Bose-Einstein distribution. For near-infrared photons with central angular frequency $\bar{\omega}=\unit[2\pi\cdot242.6]{THz}$ ($\bar\lambda= \unit[1236]{nm}$) at room temperature, the thermal occupation 
$ n_T(\bar{\omega})\approx 10^{-17}$ is negligible. 
However, one has to keep in mind that the QD-SLD is a driven semiconductor system so that the photon temperature does not have to agree with the ambient temperature. 

Commonly, the stationary field intensity 
\footnote{A common definition of an ``intensity`` misses the appropriate factor of $2 \epsilon_0 c$ \cite{tannoudjiphotons} in disagreement with the radiometric definition of intensity $\unitfrac[]{W}{m^2}$ \cite{gross2005vol1}}
of the radiation in units of $\unitfrac[]{W}{m^2}$ is given by \cite{loudon2000quantum}
\begin{align}
  I(x,y)= 2\epsilon_0 c\ave{\hat E^{(-)}(t,\vec{r})\hat E^{(+)}(t,\vec{r})}.
\end{align}
Due to the stationarity of the state and the translational invariance of the traveling wave field  in Eq.~(\ref{1.1}), the intensity is also independent of $t$ and $z$.
The optical power $P$ recorded by a 
typical single-photon detector at position $z$ is proportional to the intensity, integrated over the detector area 
\begin{align}
 \label{power}
  P&=\int_A \text{d}x\text{d}y \,I(x,y)=
	\sum_{i=1}^{N} p_i^s+p_i^t\equiv P^s+P^t
	\quad \text{with} \\
 \label{powers}
 p_i^s&=p^s(\omega_i)\equiv\frac{\hbar\omega_i c}{L}|\gamma_i|^2,\quad p_i^t=p^t(\omega_i)\equiv \frac{\hbar \omega_i c}{L}
  n_T(\omega_i).
\end{align}

The power is distributed over a bandwidth of frequencies as shown in Fig.~\ref{spectrum}.
Therefore, it is relevant to define frequency averages and variances
\begin{align}
\label{1.16}
\avee{p}\equiv \frac{1}{N}\sum_{i=1}^{N} p_i,\qquad
\Delta^2 p\equiv\sum_{i=1}^{N} 
\frac{(p_i-\avee{p})^2}{N}.
\end{align}
Consequently, the total power \eqref{power} can be expressed in terms of the average values as
\begin{align}
  \label{power2}
  P=P^s+P^t=N\left(\avee{p^s}+\avee{p^t}\right),
\end{align}
given by the sum of the average powers 
of the incoherent field $\avee{p^{s}}$ 
as well as the thermal field $\avee{p^{t}}$ times the number of modes $N$.

The physical quantities, introduced in this section, become important in the following when studying first and second order correlation functions, providing information about spectra and photon statistics of the considered light states. 
\subsection{First order temporal correlations}
According to Glauber's coherence theory 
\cite{RGlauber1963,MandelWolf}, the first order correlation function is defined as the expectation value
\begin{align}
  G^{(1)}(x_1,x_2)&=
  \ave{\hat E^{(-)}(x_1)\hat E^{(+)}(x_2)}
\end{align}
with space-time event $x=(t,\vec r)$. 
To assess scale invariant properties of the correlations, one considers normalized correlation functions usually given by the fraction
\begin{align}
  \label{1.9}
  g^{(1)}(x_1,x_2)&=
  \frac{G^{(1)}(x_1,x_2)}{\sqrt{G^{(1)}(x_1,x_1){G^{(1)}(x_2,x_2)}}}.
\end{align}
Using a spectrum analyzer, we can obtain an experimentally accessible signal that is 
proportional to the spatially averaged temporal correlation function 
\begin{align}
  \label{G1tau1}
  G^{(1)}(\tau)&=\int_A dx dy~ G^{(1)}(t,\vec r;t+\tau,\vec r).
\end{align}
Applying the normalization condition 
\eqref{normalization} and using the moments defined in Eq.~\eqref{ave}, we obtain for the temporal  first order correlation function
\begin{align}
  \label{G1tau}
  G^{(1)}(\tau)
  &=\frac{1}{2\epsilon_0 c}\sum_{i=1}^{N} e^{-i\omega_i\tau}(p_i^s+p_i^t).
\end{align}
For vanishing time delay $\tau=0$, 
the first order correlation function 
reduces to $G^{(1)}(\tau=0)=P/2\epsilon_0 c$.

In evaluating the spatially averaged, normalized first order temporal correlation function at equal position, we assume that for two different space-time events $G^{(1)}(x_1,x_2)$ changes slowly compared to equal events $G^{(1)}(x,x)$ and therefore it can be approximated by  
\begin{equation}
  \label{g1tau}
  g^{(1)}(\tau)\simeq\frac{2\epsilon_0 c G^{(1)}(\tau)}{P}=\frac{1}{P}\sum_{i=1}^{N} e^{-i\omega_i\tau}(p_i^s+p_i^t).
\end{equation}
Its modulus fulfills a Cauchy-Schwarz inequality
\begin{align}
  \label{g1lim}
  0\le|g^{(1)}(\tau)|\le|g^{(1)}(0)|=1.
\end{align}

In the experiments we evaluate field correlation spectra at the position $\vec r$, which are defined in the stationary limit as \cite{mollow69,meyestre1990} 
\begin{align}
  \label{1.6a}
  \mathcal{S}(\vec{r},\omega)&=
	\lim_{t\to\infty}\frac{\epsilon_0 c}{\pi}\int_{-\infty}^\infty d\tau~ e^{i\omega\tau}G^{(1)}(t,\vec r;t+\tau,\vec r).
\end{align}
From this definition we derive by integration over the cross section of the detector area, the power spectrum at the detector position $z$
\begin{align}
  \label{1.6b}
  S(\omega)&=
	\int_A dxdy~\mathcal{S}(\vec r,\omega) 
  =\frac{1}{\Delta\omega}\left(p^s(\omega)+p^t(\omega)\right),
\end{align}
with continuous expressions of the powers described by Eq.~\eqref{powers}. In the derivation of this result, we have approximated the sum in \eqref{G1tau} 
by the first term of the Euler-Maclaurin series (cf. Eq.~\eqref{Euler}) using the frequency separation between adjacent modes 
$\Delta\omega=(\omega_{N}-\omega_1)/(N-1)$. 
Furthermore, we have also assumed that the frequency spectrum has a finite support in the frequency band 
$[\omega_1,\omega_{N}]$ 
and the spectral width is much less than this 
bandwidth, \ie $\sigma \ll |\omega_{N}-\omega_1|$.

Obviously, the spectrum is also position independent and it consists of a superposition of the continuous distribution $|\gamma(\omega)|^2$ as well as a thermal occupation number $n_T(\omega)$. These shapes can be extracted from the measured power spectrum (see Fig.~\ref{spectrum}).

Integration of the frequency spectrum over the bandwidth 
\begin{align}
\int_{\omega_1}^{\omega_{N}} \text{d}
\omega~ S(\omega)=P
\end{align}
adds up to the total power in Eq.~\eqref{power}.

\subsection{Second order temporal correlations}
In general, two-photon correlations can be measured by two single-photon detectors \cite{BrownTwiss1956}, or a single two-photon detector \cite{mollow68}. The present experiments realize a two-photon measurement with a two-photon detector at position $z$.
The relevant observable, the second order correlation function, is defined as
\begin{align}
  G^{(2)}(x_1,x_2)=\ave{\hat E^{(-)}(x_1)\hat E^{(-)}(x_2)\hat E^{(+)}(x_2)\hat E^{(+)}(x_1)}
\end{align}
and the normalized correlation can be written as 
\begin{align}
  \label{g2}
  &g^{(2)}(x_1,x_2)=
  \frac{G^{(2)}(x_1,x_2)}{G^{(1)}(x_1,x_1){G^{(1)}(x_2,x_2)}}.
\end{align}
For slowly varying $G^{(2)}(x_1,x_2)$ 
compared to $G^{(1)}(x_1,x_1)$, the normalized temporal second order correlation function, measured by the two-photon detector, reads
\begin{align}
  \label{g2tauallg}
  g^{(2)}(\tau)\simeq \frac{(2\epsilon_0 c)^2}{P^2}\int_A dx dy~ G^{(2)}(t,\vec r;t+\tau,\vec r).
\end{align}
Evaluating the spatial integral leads to the expression 
\begin{align}
  g^{(2)}&(\tau)
 \label{g2vontau} 
  =1+|g^{(1)}(\tau)|^2-
	\frac{1}{P^2}\sum_{i=1}^{N} {{p_i}^s}^2
\end{align}
depending on the modulus of the temporal first  order correlation function $g^{(1)}(\tau)$ calculated in Eq.~\eqref{g1tau}.

For the considered PRAG state, we find that 
$g^{(2)}(\tau)$ is bounded from below and above by
\begin{align}
  0\le g^{(2)}(\tau)\le2,
\end{align}
which can be verified by considering the single terms in Eq.~\eqref{g2vontau}: the last term takes values only between 0 and 1 and the modulus of $g^{(1)}(\tau)$ is limited by \eqref{g1lim}. Furthermore, the normalized second order correlation function obeys the inequalities
\begin{align}
  g^{(2)}(\tau)\le g^{(2)}(0),\qquad g^{(2)}(0)\ge 1,
\end{align}
which also holds in the special case of treating the electrical field purely classically.

In the special case of temporal second order auto-correlation function at vanishing time difference $\tau=0$, equation \eqref{g2vontau} reduces to
\begin{align}
  \label{g20}
  g^{(2)}(0)=
  2-\frac{1}{N}\frac{1+\frac{\Delta^2p^s}{\avee{p^s}^2}}{\left(1+\frac{\avee{p^t}}{\avee{p^s}}\right)^2},
\end{align} 
with mean values $\avee{p^s},\avee{p^t}$ and variance 
$\Delta^2 p^s$ already introduced in Eq.~\eqref{1.16}.
It is interesting to note that the photon statistics of the PRAG state depends on the number of modes and their distribution, \ie $g^{(2)}(\tau)$ is coined  by the characteristics of each individual QD-SLD.

For negligible thermal contribution and limiting the 
electric field to a single mode, $N=1$, the intensity correlations
$g^{(2)}(0)=1$ are Poissonian again as demonstrated in Eq.~\eqref{poisson}.
In the complementary case of a perfect thermal light source, the second term in \eqref{g20} vanishes and consequently $g^{(2)}(0)=2$. 

%

\section{Tuning mode numbers via optical feedback}
\label{opticalfeedback}
On the one hand, the number of active modes $N$ in the emission spectrum of the QD-SLD, represents a significant parameter for the PRAG state (Eq.~\eqref{statesld}). 
On the other hand, the contribution of thermal photons 
in the near-infrared $n_T(\omega)$ is marginal for room temperatures and will be neglected in the following. The inverse proportionality to $N$ in Eq.~\eqref{g20} suggests, that for a high number of modes, the intensity correlations should be very close to 
$g^{(2)}(0)=2$, whereas for a low number of modes, 
$g^{(2)}(0)\rightarrow 1$ continuously.

\begin{figure}[h]
   \includegraphics[width=\columnwidth]{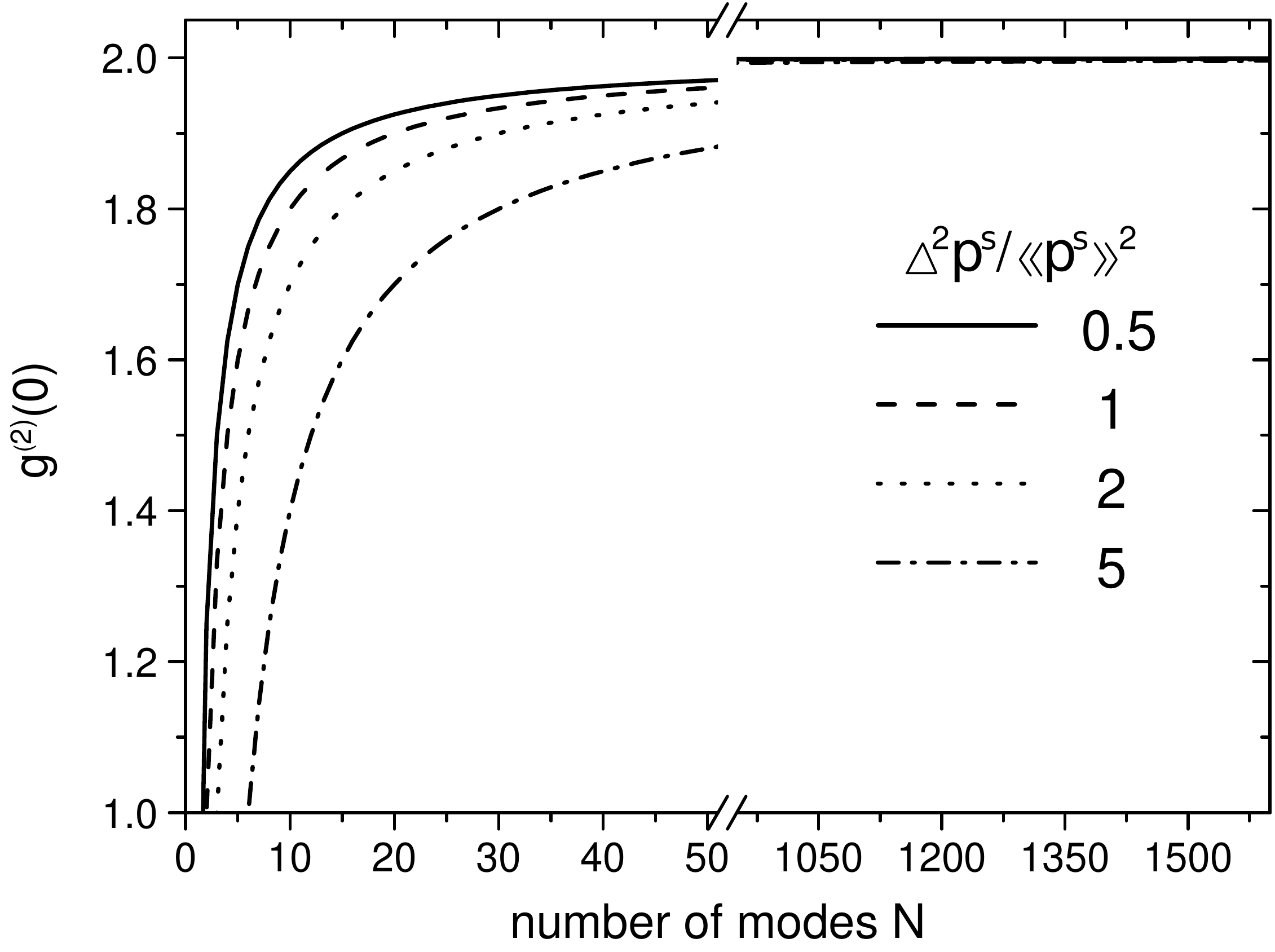}
    \caption{Intensity correlations $g^{(2)}(0)$
		versus number of modes $N$. 
		We show the influence of the relative 
		spectral width $\Delta^2 p^s/\avee{p^s}^2$ for a PRAG state.}
  \label{numofmodes}
\end{figure}

Fig.~\ref{numofmodes} visualizes the dependence
 of $g^{(2)}(0)$ as a function of $N$ for different values of $\Delta^2 p^s/\avee{p^s}^2$: they all show steep trajectories from $g^{(2)}(0)=1$ to $g^{(2)}(0)=2$, where with increasing ratio 
 $\Delta^2 p^s/\avee{p^s}^2$, $g^{(2)}(0)$ functions are shifted towards higher values of $N$.

\begin{figure*}
   \includegraphics[width=1.8\columnwidth]{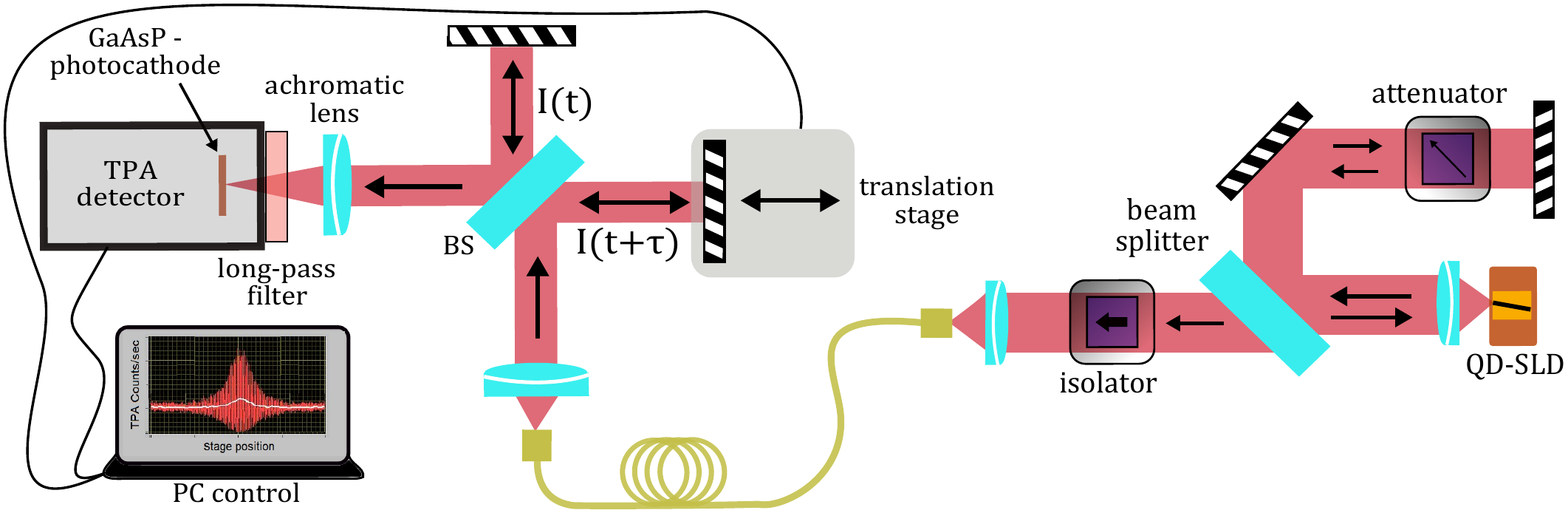}
    \caption{(Color online) (Left) Schematics of the TPA-detection setup for second order correlation measurement \cite{Boitier2009}. (Right) Schematics of the optical feedback experiment.}
  \label{TPAsetup}
\end{figure*}

We put these theoretical predictions to an experimental trial with a QD-SLD. 
Therefore, we must measure the second order correlation functions $g^{(2)}(\tau)$ of light emitted in the near-infrared with spectral widths up to more than $\Delta\lambda=\unit[100]{nm}$ corresponding to $b=\unit[2\pi\cdot19.9]{THz}$ in terms of angular frequency, which sets challenging requirements to the time resolution of the measurement system. In 2009, Boitier and coworkers \cite{Boitier2009}
developed a method to experimentally access sub-femtosecond time-resolution for second order correlation functions $g^{(2)}(\tau)$. 
The technique is based on two-photon-absorption (TPA) inside a semiconductor-based photocathode of a photomultiplier tube (PMT). 
TPA is an absorption process, which relies on a virtual state inside the bandgap of the semiconductor material exhibiting a lifetime resulting from energy-time uncertainty, thus enabling ultrafast and broadband detection of the expectation value of $\hat E^{(-)}(t)\hat E^{(-)}(t)\hat E^{(+)}(t)\hat E^{(+)}(t)$
\cite{mollow68}. Implementing the TPA-PMT in a Michelson-Interferometer which introduces a time delay $\tau$ via a high precision, motorized translation stage, second order autocorrelation functions $G^{(2)}(\tau)
$ 
can be extracted from the measured TPA-interferograms via low-pass filtering (Fig.~\ref{TPAsetup} (left)) \cite{Boitier2009,Mogi1988}.

In the following, we will use the notation 
$g_{\text{th}}^{(2)}(0)$ and $g_{\text{exp}}^{(2)}(0)$ 
in order to differ between theoretically predicted and experimentally determined 
$g^{(2)}(0)$ values, respectively.
It turns out that one of our earlier studies 
\cite{Hartmann2013} demonstrates the tailoring 
of first and second  order coherence properties of pure QD-SLD emission by applying optical 
feedback (OFB) onto the semiconductor emitter 
(Fig.~\ref{TPAsetup} (right)).
The essence of this investigation was the observation of a simultaneous, continuous reduction of i) the spectral width $\Delta\lambda$ from 120nm to  sub-nanometer values and ii) the second order coherence degree $g_{\text{exp}}^{(2)}(0)$ from 1.85 to 1.0, for the light emitted by the QD-SLD (InAs/InGaAs - Dot in Well - structure) under increased OFB. 
However, this observed transition in coherence, induced at relatively low spectral widths, still lacks a theoretical explanation. 
This published experimental investigation is therefore perfectly suited to be compared to the here performed theoretical investigation, especially because narrowing the spectral width is synonymous with reducing the number of modes $N$.
The spectral shapes of the different emission regimes induced by the OFB, range from ultrabroadband ASE to multimode emission where only few modes appear. Especially the multimode emission regime allows to enumerate specifically the number of modes $N$ contributing to the global light state. 
\begin{figure}[h!]
   \includegraphics[width=\columnwidth]{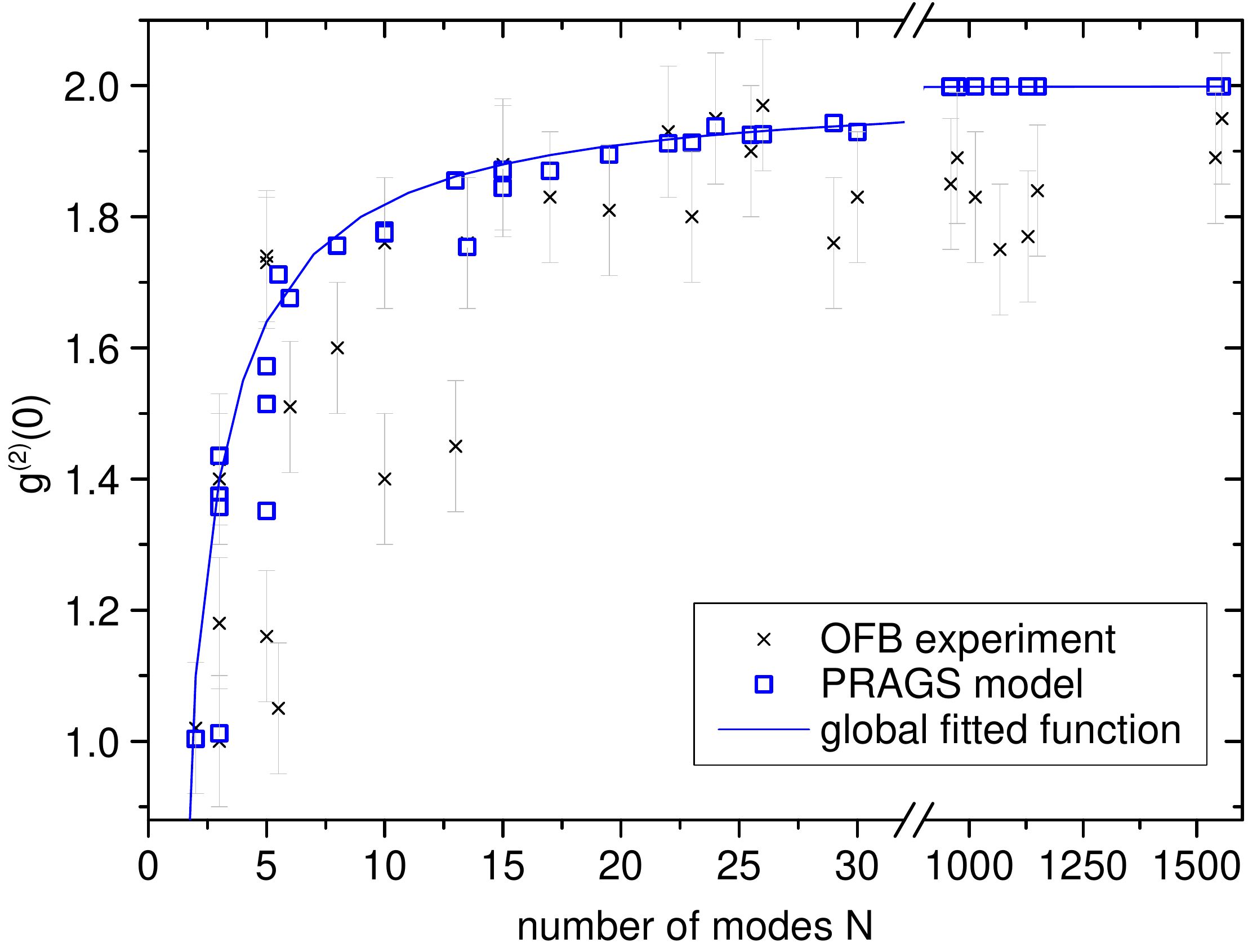}
    \caption{Results of the second order coherence degree $g^{(2)}(0)$ from the optical feedback experiment as a function of the number of modes $N$ in
		comparison with values calculated according to our theoretical model. A global fit with a relative spectral width of 0.8 is depicted as a guide-to-the-eye.
		The depicted error bars visualize the 
		representative uncertainty of one shot 
		measurements which were imposed due to 
		challenging stabilization conditions of the QD-
		SLD under OFB.}
  \label{feedback}
\end{figure}

The continuous transition of the second order coherence degree $g_{\text{exp}}^{(2)}(0)$ taken from 
\cite{Hartmann2013}, is now depicted in 
Fig.~\ref{feedback} as a function of the number of emitted modes $N$, calculated from the measured optical spectra $p^s_i$. 
One should note that it is essential to exclude nonrelevant spectral contributions, which can falsify the statistics of $p^s_i$, thus we choose to take into account only those peaks, not being more than 
$\unit[13]{dB}$ below the maximum power value 
$p^s_{\text{max}}$.
We calculate the corresponding theoretical values 
$g_{\text{th}}^{(2)}(0)$ according to Eq.~\eqref{g20} with
 the experimentally obtained parameters $N$, 
$\avee{p^s}$ and $\Delta^2 p^s$ in order 
to reproduce the experimental conditions of the observed coherence transition. 
Fig.~\ref{feedback} matches experimental data with theoretical prediction. Numerical values are tabulated in Tab.~\ref{Tab1}, for reference.

For ultrabroadband QD-SLD emission, $N$ takes very 
high values. Here, the number of modes could not be
 enumerated straightforwardly by counting spectral 
peaks, because smooth Gaussian-like spectral shapes 
dominate and therefore $N$ remains experimentally 
undeterminable. However, a lower bound estimate is 
given by the number of Fabry-Pérot modes matching the 
length $L=\unit[3]{mm}$ of the QD-SLD waveguide, similar to a 
multimode laser but here with strongly broadened as 
well as overlapping longitudinal modes. In practice, 
$N$ has been determined by fitting modes with spacing 
according to the free spectral range (FSR)
in terms of angular frequency 
$\Delta\omega=2\pi c/2n_
{GaAs}L=\unit[2\pi\cdot 1.465\cdot10^{10}]{Hz}$ ($n_{GaAs}\approx3.41$) 
to the optical spectra taking into account the above 
mentioned $\unit[13]{dB}$ cutoff, resulting in mode numbers $N>
1000$. 
In this regime, we observe experimental values 
$g_{\text{exp}}^{(2)}(0)$ fluctuating around 1.85 and 
theoretical values around $g_{\text{th}}^{(2)}(0)=1.999$,
\ie very close to the limit value of 2
for pure thermal states. The discrepancy 
$\Delta g^{(2)}(0)=|g_{\text{exp}}^{
(2)}(0)$–$g_{\text{th}}^{(2)}(0)|\approx 0.15$ 
between 
experiment and theory in 
this ultrabroadband regime, can be attributed to the 
frequency-dependency of the TPA absorption parameter $
\beta(\omega)$ of the detecting photomultiplier 
\cite{Boitier}, 
which is not able to provide ideal equal 
detection efficiency over the total range of 
frequencies. Again, the here specified mode numbers $N$
 are lower bound estimates, but regarding 
Fig.~\ref{numofmodes}, 
we can deduce that, considering the 
experimentally determined values of $\Delta^2 p^s/
\avee{p^s}^2$ (see Tab.~\ref{Tab1}) of about 
0.6, $g_{\text{th}}^{(2)}(0)(N>1000)$ is clearly restricted 
to values above 1.99, which limits the uncertainties $
\Delta g_{\text{th}}^{(2)}(0)$ to below 1\%.
\begin{table}[h!]
\begin{ruledtabular}
    \begin{tabular}{ c  c  c  c }
    $N$ & $\frac{\Delta^2 p^s}{\avee{p^s}^2}$ & $g^{(2)}_{\text{exp}}(0)$ & $g^{(2)}_{\text{th}}(0)$\\ \hline 
    3 & 1.31 & 1.18 & 1.23 \\ \hline
    10 & 1.12 & 1.78 & 1.74\\ \hline
    30 & 1.08 & 1.83 & 1.931\\ \hline
    1945 & 0.57 & 1.84 & 1.999 \\
    \end{tabular}
		\end{ruledtabular}
    \caption{Representative values taken from Fig.~\ref{feedback} with experimentally determined parameters $N$ and $\Delta^2 p^s/\avee{p^s}^2$ for calculation using Eq.~\eqref{g20}.}
    \label{Tab1}
\end{table}

Entering the regime of directly countable mode numbers of $N=30$ down to $N=15$, we still observe high second order coherence degrees above $g_{\text{exp}}^{(2)}(0)=1.8$, however already with a slightly decreasing tendency. This is in agreement with the calculated values $g_{\text{th}}^{(2)}(0)$ which show a less fluctuating trajectory. It is only for small mode numbers $N<15$, that a steep transition from $g^{(2)}(0)=1.8$ to $g^{(2)}(0)=1.0$ is recorded, both for experimental as well as for calculated values. Strongly deviating $g_{\text{exp}}^{(2)}(0)$ values are due to challenging experimental conditions concerning the stabilization of the QD-SLD emission under OFB during the measurement. Nevertheless, the agreement between experiment and theory is more than obvious and therefore we can confirm, that the coherence transition is indeed triggered by the strongly reduced number of existing emission modes N and the slightly enhanced ratio of  
$\Delta^2 p^s/\avee{p^s}^2$. Hence, this result supports the assumed PRAG state for describing ASE light states from QD-SLDs.

Unfortunately, the coherence transition is observed for very low number of modes where the QD-SLD does no longer exhibit smooth broadband spectra. The 
reason for significant second  order coherence changes only for $N<15$, lies in the small values of $\Delta^2 p^s/\avee{p^s}^2$ 
(see Tab.~\ref{Tab1}) in the range between 1 and 2. For broadband emission with tens of nanometer spectral widths and Gaussian-like spectral shapes, we find even lower values  $\Delta^2 p^s/\avee{p^s}^2<1$, which fix second order coherence degrees quickly to $g^{(2)}(0)=2$ by increasing $N$ (Fig.~\ref{numofmodes}). 
The drawback of these results is therefore the loss
 of the broadband emission property of the QD-SLD and 
hence, the accuracy of the PRAG model in the broadband ASE 
regime of the QD-SLD still requires more evidence.

Consequently, we choose to implement a second experimental 
approach with priority to the conservation of the 
broadband ASE regime of QD-SLD operation: a fully 
coherent light state from single mode laser emission 
is superimposed to broadband ASE from a QD-SLD with an 
implemented variability of the intensity ratio 
between both light components influencing the second 
order correlation properties. 
The coherent light state thereby probes the accuracy of the assumed PRAG state via the combined photon statistical behavior. 
This approach is based on the concept of “mixed 
light”, which has been subject to extensive 
experimental and theoretical studies, starting 
shortly after the invention of the laser in the 1960s 
in connection with photon counting methods and the 
newly developed Hanburry-Brown Twiss experiment 
\cite{Arecchi1965,Arecchi1966,glauberlectures1965,Scarl1972}. 
Recently, mixed light state analysis with pseudothermal light \cite{Martienssen1964} 
has been investigated, demonstrating the continuous
 tunability of photon statistics \cite{Lee2011} also 
regarding polarization dependencies related to 
possible applications such as ghost imaging schemes 
\cite{Liu2014}. Here, we extend the mixed light 
phenomenon to highly first order incoherent light 
sources and we moreover exploit it for the 
verification of our theoretical model.

\section{Mixing light from two sources}
\label{heterodyne}
In this section, we present the theoretical analysis of the superposition of a coherent light state with the already introduced PRAG state, focusing on the proper quantum optical definition of the superimposed state of light and the resulting combined second order correlation behavior. In a second step we will show results of the realization of a mixed light experiment.

\subsection{Mixing light theoretically}
\begin{figure}[h]
    \includegraphics[width=\columnwidth]{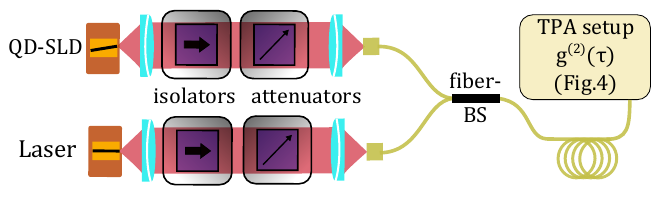}
    \caption{Schematic of the setup for the mixed light experiment with all semiconductor-based free space emitting devices. For second order correlation analysis, the fibercoupled mixed light is guided to the TPA-detection setup in Fig.~\ref{TPAsetup}}
    \label{mixedlightsetup}
\end{figure}  
According to the here implemented mixed light experiment (see Fig.~\ref{mixedlightsetup}), light from a QD-SLD is superimposed 
with light generated by an independent single-mode laser source with a fixed frequency $\omega_k$ combined in a fiber-based beam splitter. From there on, the state of the electric field reads 
\begin{eqnarray}
 \label{2.1}
 \hat \rho_m=\hat D(\alpha_k)\hat \rho_{s}\hat D^\dagger(\alpha_k),
\end{eqnarray}
with the single-mode displacement operator $\hat{D}(\alpha_k)=\text{exp}(\alpha_k^{ }\hat{a}_k^\dagger-\alpha_k^\ast\hat{a}_k^{ })$ and 
$\hat{\rho}_s$ given by Eq.~\eqref{statesld}. 
In other words, we add a coherent amplitude 
$\alpha_k$ in mode $k$ to the state of the 
QD-SLD light as a result of the beam splitter, mixing the two independent sources.

\begin{figure}[h!]
    \centering
     \scalebox{1}{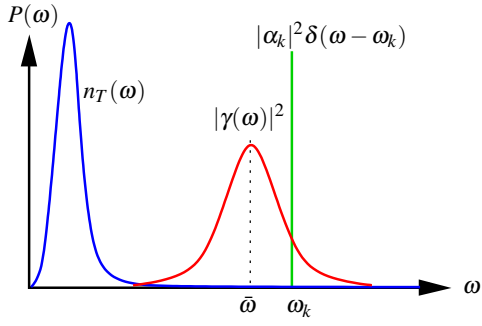}
    \caption{Schematic representation of the single components of the spectral distribution for mixed light. The depicted curves are: a delta distribution (green) for the laser light, a Gaussian distribution (red) for the QD-SLD light and a thermal Planck distribution (blue), for reference.}
  \label{distribution}
\end{figure}

The normalized temporal first and second order autocorrelation functions for mixed light can be determined in the same way as in the case of a single source.
For the temporal second  order correlation function $G^{(1)}(\tau)$ and the total power $P^m$ of light states characterized by the density operator of Eq.~\eqref{2.1} one gets
\begin{align}
  \label{G1tauallg}
  G^{(1)}(\tau)&=
  \sum_{i=1}^{N} 
	e^{-i\omega_i\tau}(p_i^l+p_i^s+p_i^t),\\
  P^m&=\sum_{i=1}^{N} p^l_i+p_i^s+p_i^t\equiv P^l+P^s+P^t,
\end{align}
showing the same results as for the QD-SLD but with additional terms considering contributions from the laser. Here, the laser power is defined as
\begin{align}
  P^l\equiv \sum_{i=1}^{N}p^l_i \quad \text{with} \quad p^l_i=\frac{\hbar\omega_i c}{L}|\alpha_i|^2\delta_{ik}.
\end{align}
Now, we can specify the spectrum of the mixed light state,
\begin{align}
 \label{spectrummixed}
  S(\omega)&=
	P^l\delta(\omega-\omega_k)+\frac{1}{\Delta\omega}\left(p^s(\omega)+p^t(\omega)\right),
\end{align}
with three contributing terms, \ie three single spectral distributions, as illustrated in 
Fig.~\ref{distribution}.

The green line indicates a delta function at frequency $\omega_k$, occurring due to the laser light description of a pure coherent state. The other two distributions originate from the assumed nature of the PRAG states: the blue curve reflects the thermal contribution, described by an ordinary Planck distribution and the red one is a Gaussian, representing its incoherent character.

The temporal normalized second order correlation function in the case of mixed light with density operator $\hat{\rho}_m$ reads 
\begin{align}
  \label{g2vontauallg}
  g^{(2)}(\tau)
  =1+|g^{(1)}(\tau)|^2-\frac{{P^l}^2+
	\sum_{i=1}^{N} {p_i^s}^2}{{P^m}^2}. 
\end{align}  
As in the previous discussion of Sec.~II, the time-dependence only arises from the modulus of
\begin{align}
	\label{G1square}
  |g^{(1)}(\tau)|^2&=
  \frac{1}{{P^m}^2}\Bigl[ 
  {P^l}^2+|\sum_{i=1}^{N} 
	p_i^se^{-i\omega_i\tau}|^2+|\sum_{i=1}^{N} 
	p_i^te^{-i\omega_i\tau}|^2 \\ \nonumber
  &+2\sum_{i,j=1}^{N} p_j^s p_i^t
	\cos{\Delta_{ij}\tau}+
	2P^l\sum_{i=1}^{N} (p_i^s+p_i^t)
	\cos{\Delta_{ik}\tau}\Bigr]
\end{align}
with frequency difference 
$\Delta_{ij}=\omega_i-\omega_j$. 
The last term in 
\eqref{G1square} oscillates with the beat frequency 
of the laser and the $i$th mode of the QD-SLD $\Delta_{ik}=\omega_i-\omega_k$, leading to side bands in the spectrum and finally in $g^{(2)}(\tau)$. 

It is instructive to discuss different limiting cases of the light state assumptions. For a pure thermal state, \ie $\alpha_k$ and $\gamma_i$ are zero, the last term in \eqref{g2vontauallg} vanishes and the temporal second order correlation function reduces to the well known Siegert relation \cite{jahnke2012}
\begin{align}
  g^{(2)}(\tau)=1+|g^{(1)}(\tau)|^2,
\end{align}
where $g^{(1)}(\tau)$ is the normalized first order correlation function for thermal light sources. 
For a single coherent state, the correlation function of second order takes the expected constant value of one, $g^{(2)}(\tau)=1$, for arbitrary time delay $\tau$. 
Certainly, for vanishing amplitude $\alpha_k$ the expression of second order correlation function, as already studied, reduces to Eq.~\eqref{g20}.
Especially, in the case of identical space-time events, $\tau=0$, we get 
\begin{align}
  \label{g20mixedlight}
  g^{(2)}(0)
  =2-\frac{{P^l}^2+\sum_{i=1}^{N} {p_i^s}^2}{{P^m}^2}.
\end{align}
Rewriting $g^{(2)}(0)$ in terms of variance and mean values, already introduced, yields
\begin{align}
  \label{g20allg}
  g^{(2)}(0)&=2-\frac{1}{N}
	\frac{1+\frac{\Delta^2 p^s}{\avee{p^s}^2}
  +\frac{{P^l}^2}{N\avee{p^s}^2}}{\left(
	1
	+\frac{\avee{p^t}}{\avee{p^s}}+
	\frac{P^l}{N\avee{p^s}}\right)^2}.
\end{align}

\subsection{Example of a Gaussian spectrum}
Motivated by the experimentally obtained optical spectra of Fig.~\ref{spectrum}, we study analytically the case of a single Gaussian spectrum, \ie
\begin{eqnarray}
 \label{gauss}
  p^s(\omega)=P^s_0\frac{\omega\Delta\omega}{\sqrt{2\pi}\sigma\bar\omega}e^{-\frac{(\omega-\bar\omega)^2}{2\sigma^2}},
\end{eqnarray}
with mean value $\bar\omega$, frequency width $\Delta\omega=(\omega_{N}-\omega_1)/(N-1)$ and standard deviation $\sigma$. 
The normalization constant $P^s_0$ is determined by 
the discrete summation of the powers
\begin{align}
P^s\equiv\sum_{i=1}^{N} p^s(\omega_i)\approx P^s_0,
\end{align}
which is satisfied by Eq.~\eqref{gauss} assuming the applicability of the Euler-Maclaurin formula (see Appendix B). For the sake of simplicity we neglect thermal contributions to the spectrum of the SLD, \ie $p^t(\omega_i)=0$.
After utilization of the Euler-Maclaurin formula and introduction of dimensionless variables $\tilde\tau=\sigma\tau$, $\tilde{\bar\omega}=\bar\omega/\sigma$, $\Delta\tilde\omega=\Delta\omega/\sigma$, $\delta\tilde\omega_k=\delta\omega_k/\sigma=(\bar\omega-\omega_k)/\sigma$ we obtain the scaled second order correlation function 
\begin{align}
  \label{g2gaussian}
  g^{(2)}(\tilde\tau)&=
	1+\frac{1}{(1+\epsilon)^2}
	\times
	\{
	e^{-\tilde\tau^2}
	(
		1+(\tilde\tau/\tilde{\bar\omega})^2)-\eta \\ \nonumber
  &+2e^{-\frac{\tilde\tau^2}{2}}
	\epsilon
	[\cos{(\delta\tilde\omega_k \tilde\tau)}
	-\frac{\tilde\tau}{\tilde{\bar\omega}}\sin{(\delta\tilde{\omega}_k\tilde\tau})]
 \},\\
 &\text{with}\qquad
 \label{epsilon}
\eta=\frac{\Delta\tilde\omega}{2\sqrt{\pi}}\left(1+\frac{1}{2\tilde{\bar\omega}^2}\right),\qquad
\epsilon=\frac{P^l}{P^s}.
\end{align}

\begin{figure}[h!]
    \includegraphics[width=.8\columnwidth]{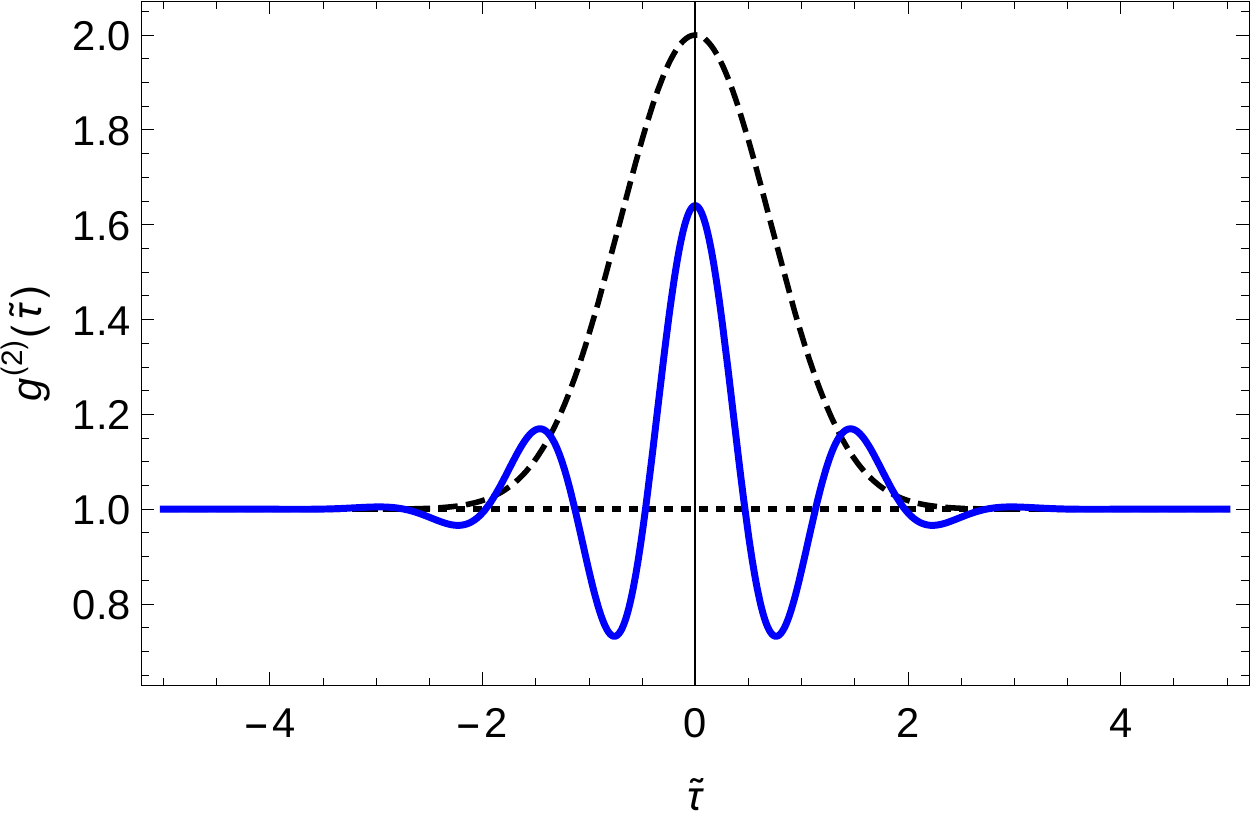}
     \caption{Scaled temporal second  order correlation function $g^{(2)}(\tilde{\tau})$ versus dimensionless delay time $\tilde{\tau}$ (Eq.~\eqref{g2gaussian}) for Gaussian distributed photon number $|\gamma_i|^2$ with $\epsilon=3/2$,  $\Delta\tilde\omega=10^{-3}$, $\tilde{\bar\omega}=100$ and $\delta\tilde\omega_k=4$ (blue line). The black dotted (dashed) line represents the correlation in absence of the light emitting QD-SLD (single-mode laser).}
   \label{g2gauss}
 \end{figure}

For increasing time delay, the first term in the brackets exhibits an exponential decreasing behavior which is subtracted by a small offset depending on the frequency distance $\Delta\tilde\omega$ as well as the mean value $\tilde{\bar\omega}$ of the QD-SLD and the last term shows a damped oscillation with beat frequency $\delta\tilde\omega_k$, depicted in Fig.~\ref{g2gauss}.
Here, the blue line corresponds to the scaled second order correlation function of mixed light for varying time delay $\bar\tau$ with the chosen values $\epsilon=3/2$, $\Delta\tilde\omega=10^{-3}$, $\tilde{\bar\omega}=100$ and $\delta\tilde\omega_k=4$. The dashed (dotted) black line depicts the limiting case of vanishing laser (QD-SLD) light.    
Evaluating Eq.~\eqref{g2gaussian} for time delay $\tau=0$ results in 
\begin{align}
  \label{g20gauss}
  g^{(2)}(0)&=
  2-\frac{\eta+{\epsilon}^2}{(1+\epsilon)^2},
\end{align}
depending on the frequency width $\Delta\omega$, $\sigma$ as well as on the ratio of the powers of the laser and the QD-SLD.

\subsection{Mixing light experimentally}

The superposition of the coherent light state with broadband QD-SLD light is experimentally realized by the exclusive use of semiconductor-based opto-electronic emitters, namely a single-mode quantum-well ridge-waveguide Fabry-Pérot laser (Eblana Photonics) and a quantum-dot superluminescent diode (Innolume GmbH).

\begin{figure}%
   \centering
   \includegraphics[width=\columnwidth]{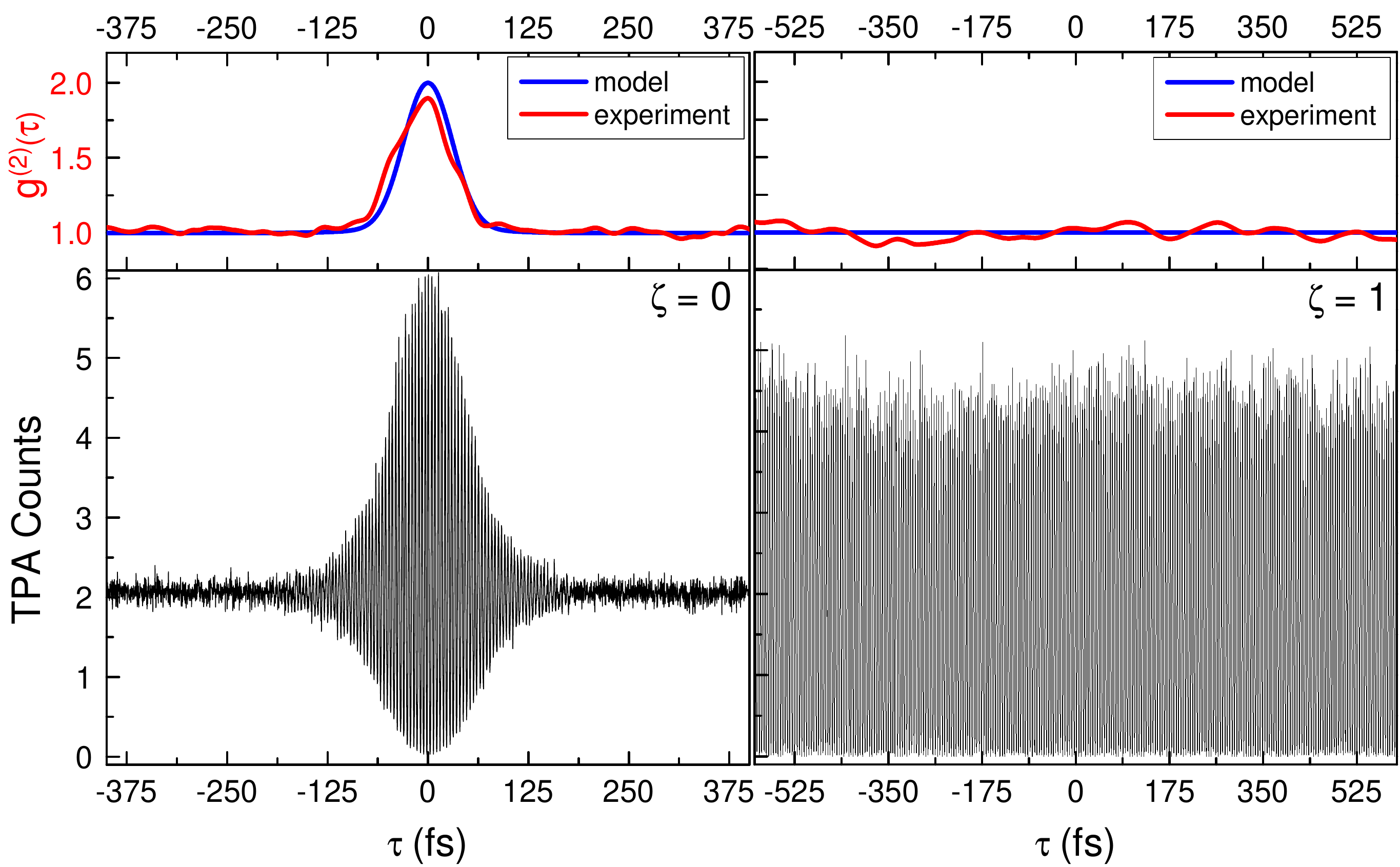}\hfill
     \caption{TPA interferograms in arbitrary TPA Counts units (bottom) and $g^{(2)}(\tau)$ (top) of single device emission: (left) QD-SLD ASE ($\zeta=0$) and (right) single-mode laser light ($\zeta=1$)}
   \label{tpa-interferogram1}
 \end{figure}

The 4mm long waveguide QD-SLD consists of a triple chirped epitaxial structure (InAs/InGaAs - Dot in Well structure with 10 active QD-layers) in order to realize ultrabroadband ASE when operated above ASE threshold with a spectral width $\Delta\lambda=\unit[67]{nm}$ centered at approximately $\bar\lambda=\unit[1236]{nm}$ (Fig.~\ref{spectrum}). The combination of a high reflective facet on the backside and an anti-reflective facet on the front side, allows i) high intensities and efficient light-outcoupling as well as ii) efficient suppression of reflections back into the waveguide at the output facet in order to prevent spectral narrowing. This extreme first order incoherence is accompanied by enhanced second order correlations within the ultrashort coherence time, visible solely on the scale of approximatively $\unit[100]{fs}$ on the recorded TPA interferogram (Fig.~\ref{tpa-interferogram1} (bottom left)).

Fig.~\ref{tpa-interferogram1} (top left) depicts the extracted second order correlation function $g^{(2)}(\tau)$ (red line) together with the theoretical counterpart (blue line), calculated according to the PRAG state model (Eq. \eqref{g2vontau}) with experimentally determined parameters taken from the corresponding optical spectra: $N$, $p^s_{i}$, $P$ and also $|g^{(1)}(\tau)|^2$ using Eq. \eqref{g1tau}. Just like for the optical feedback experiment, $N$ is estimated by taking the lower bound of possibly contributing modes, namely the number of Fabry-Pérot modes fitting into the recorded optical spectrum with spacing corresponding to the FSR with respect to the here employed $\unit[4]{mm}$ long waveguide of the QD-SLD. One can recognize well coinciding functions, both revealing i) an ultrashort coherence time of $\unit[70]{fs}$ and ii) strongly enhanced correlations with a central second order coherence degree of $g_{\text{exp}}^{(2)}(0)=1.91\pm0.05$, close to the limit value of 2 for pure thermal states, which is nicely reproduced by theory ($g_{\text{th}}^{(2)}(0)=1.999$), revealing a fully incoherent light state for the QD-SLD emission.

On the other hand, the single-mode laser, operated above laser threshold, exhibits a central wavelength of $\bar\lambda=\unit[1300]{nm}$ in combination with a spectral bandwidth $b<\unit[2\pi\cdot1.75\cdot 10^{6}]{Hz}$ in terms of angular frequency $\omega$ as well as a side-mode suppression ratio of $\unit[37]{dB}$. Ideal coherent laser light exhibits constant correlation functions $g_{\text{th}}^{(n)}(\tau)=1$ for every order $n$ and thus $g_{\text{th}}^{(2)}(\tau)=1$ is expected. Measuring the second order correlation function, delivers an approximate constant value of $g_{\text{exp}}^{(2)}(\tau)=1.01\pm0.04$ (Fig.~\ref{tpa-interferogram1} (right)), which reveals a high coherent light source character, not only showing high quality monochromaticity reflected by the fully modulated interference fringes (Fig.~\ref{tpa-interferogram1}  (bottom right)), but also a central second order coherence degree of $g_{\text{exp}}^{(2)}(0)=1.00\pm0.01$ (Fig.~\ref{tpa-interferogram1} (top right)), reflecting Poissonian photon statistics behavior \footnote{Because of the limited range of the translation stage moving the mirror inside the interferometer, this value has been double-checked via a photon-counting experiment determining the explicit photon number distribution \cite{Koczyk1996} $p(n)$, validating $g_{\text{exp}}^{(2)}(0)=1.0$}.

 The experimental setup for the superposition of the two light fields, already introduced in the theoretical part, is schematically drawn in Fig.~\ref{mixedlightsetup}. In order to get experimental access to a maximum range of photon statistical variation in terms of $g^{(2)}(0)$, we introduce a variable attenuator within the beam path of each light source.

As stated in the beginning, special care is taken to preserve the broadband QD-SLD emission property and hence we ensure steady state emission conditions by driving both light sources at a constant heat sink temperature of $\unit[20.0]{^\circ C}$ and constant DC-pump currents. The combination of the two respective attenuation values results in a power ratio between single-mode laser optical power $P^l$ and QD-SLD optical power $P^s$ which represents the critical parameter for the photon statistics tunability in the mixed light experiment (Eq. \eqref{g20mixedlight} and Eq. \eqref{g20allg}). A more clear illustration of its dependency can be given by introducing a relative quantity $\zeta$ expressed by
\begin{align}
\label{zeta}
\zeta=\frac{P^l}{P^l+P^s}=\frac{\epsilon}{1+\epsilon},
\end{align}
with $\epsilon$ already defined in Eq.~\eqref{epsilon}, constraining the values of $\zeta$ to a range between 0 (exclusive QD-SLD emission) and 1 (exclusive laser emission). Applying $\zeta$ to the theoretical results of the mixed state of light, we can rewrite Eq.~\eqref{g2vontauallg} and Eq.~\eqref{g20allg} into
\begin{align}
\label{g2vontauallgZeta}
g^{(2)}_{\text{th}}(\tau)=1+|g^{(1)}_{\text{th}}(\tau)|^2
-\zeta^2-\frac{1}{N}\left[1+\frac{\Delta^2 p^s}{\avee{p^s}^2}\right](1-\zeta)^2
\end{align}
and correspondingly
\begin{align}
\label{g20allgZeta}
g^{(2)}_{\text{th}}(0)=2-\zeta^2-\frac{1}{N}\left[1+\frac{\Delta^2 p^s}{\avee{p^s}^2}\right](1-\zeta)^2.
\end{align}
Note that these theoretical counterparts respect the general and discrete spectral distribution case $p^s_{i}$ due to complex optical spectra formation of the QD-SLD (see Fig.~\ref{spectrum}) \cite{Grundmann2002} and they will be utilized to calculate theoretical counterparts for the following comparisons to experimental results.

 \begin{figure}[h]
    \includegraphics[width=0.45\textwidth]{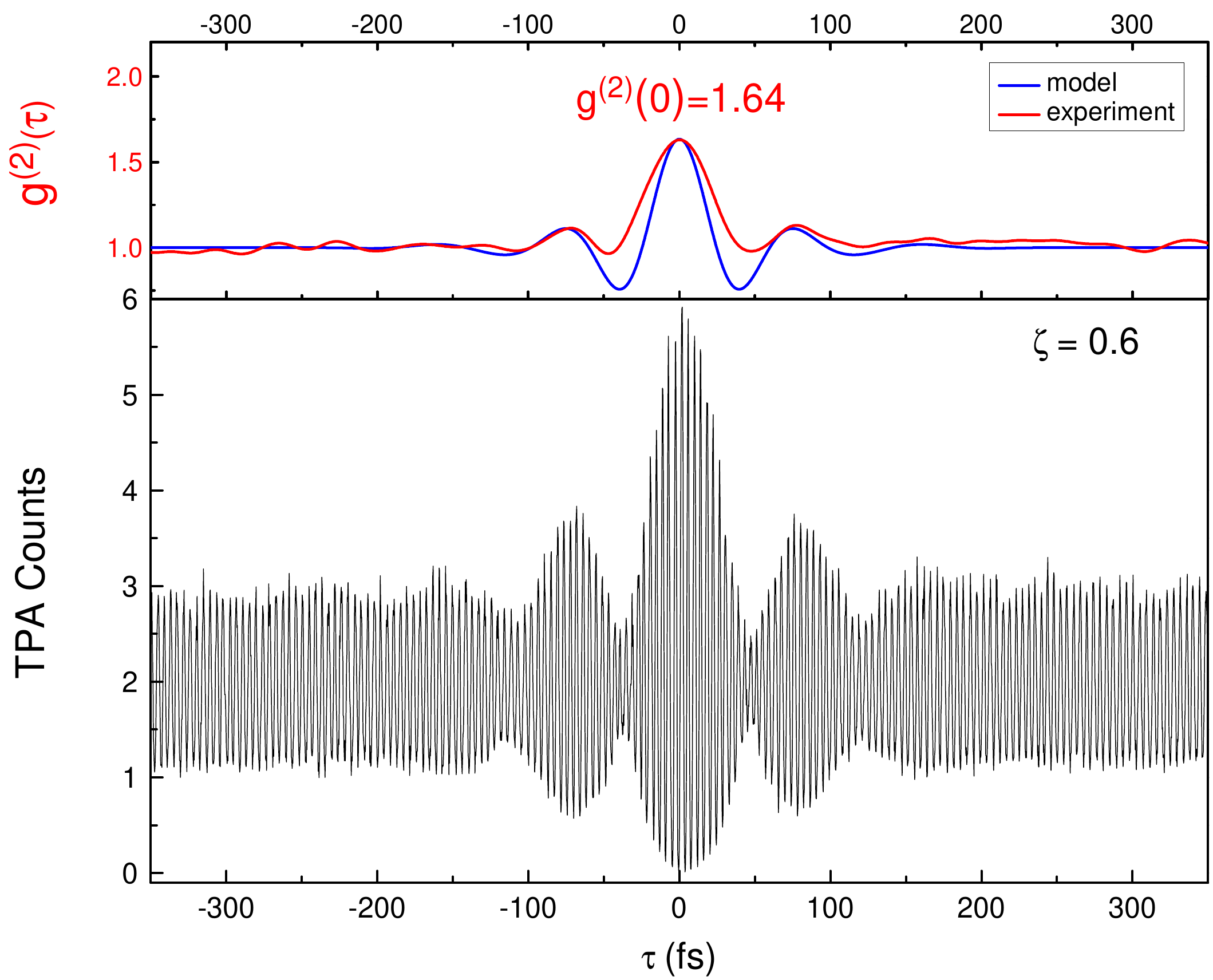}
     \caption{Exemplary TPA-Interferogram in arbitrary TPA Counts units (bottom) and $g^{(2)}(\tau)$ (top) of a mixed light realization.}
   \label{tpa_interferomixed}
 \end{figure}

Fig.~\ref{tpa_interferomixed} (bottom) shows an exemplary TPA interferogram corresponding to $\zeta=0.6$. The interferogram exhibits a shape including features from both sources: i) a long range ($\tau/\tau_c\gg1$) intensity modulation originating from laser emission, however with reduced constructive and destructive interference maxima which show already the interplay of both light fields and ii) enhanced correlation for $\tau/\tau_c<1$ originating from QD-SLD emission together with a modulation of the envelope, clearly indicating a superposition. Fig.~\ref{tpa_interferomixed} (top) pictures the experimentally extracted $g_{\text{exp}}^{(2)}(\tau)$ function (red line) as well as the calculated correlation function $g_{\text{th}}^{(2)}(\tau)$ (Eq.~\eqref{g2vontauallgZeta}, blue line) showing well coinciding trajectories: the beat signal like modulation of the envelope of the interferogram (Fig.~\ref{tpa_interferomixed} (bottom)) translates into secondary maxima $g^{(2)}(\pm\tau_2)$ (Fig.~\ref{tpa_interferomixed} (top)), corresponding to the spread of the central wavelengths of both emitters $\Delta\lambda\approx\unit[64]{nm}$, resulting in a beat time of $\tau_{\text{beat}}=\tau_2\approx\unit[76]{fs}$ where the theoretical model reproduces nicely both, the proper time scales $+\tau_2$ and -$\tau_2$ as well as the absolute values of the secondary maxima $g_{\text{th}}^{(2)}(\pm\tau_2)=1.1$. Most decisively, $g_{\text{exp}}^{(2)}(0)$ takes a value of 1.64, clearly differing from values of the two single emission states also confirmed by theory with a value 1.63. Slight deviations between theory and experiment are observed in the range of $0<\tau<\tau_2$ where the experimental resolution does not allow to record the theoretically predicted minima below $g^{(2)}(\tau)=1$. \newline
\begin{figure}
    \includegraphics[width=0.5\textwidth]{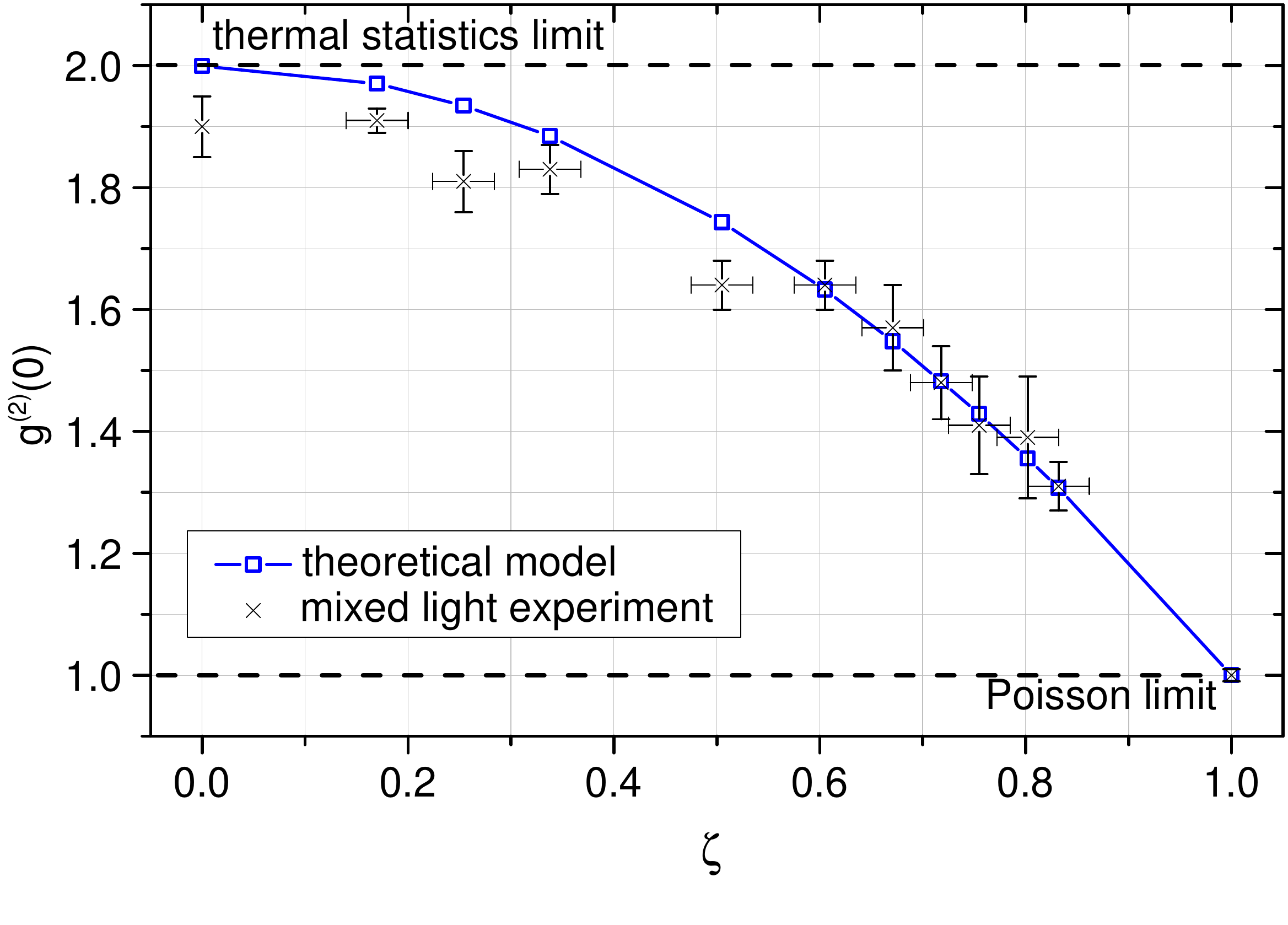}
     \caption{(Color online) Photon statistics results in terms of $g_{\text{exp}}^{(2)}(0)$ (black crosses) as a function of the relative parameter $\zeta$ together with error bars resulting from standard deviation of five averaged experimental values. Calculated values of $g_{\text{th}}^{(2)}(0)$ (blue squares) obtained from our theoretical model based on experimental parameters are also plotted for comparison.}
   \label{g2exp_theo}
 \end{figure}
Investigating $g^{(2)}(0)$ as a function of $\zeta$ (Fig.~\ref{g2exp_theo} black crosses), we achieve a full range, continuous tunability of $g^{(2)}(0)$ in the range between 1.91 and 1.0 with a parabola-like trajectory. To the best of our knowledge, this is the first demonstration of the mixed light phenomenon including an ultrabroadband light source. Fig.~\ref{g2exp_theo} also depicts the theoretical values $g^{(2)}_{\text{th}}(0)$ (blue squares), obtained from the derived analytical expression (Eq.~\eqref{g20allgZeta}) calculated with the experimentally determined parameters: $\zeta$, $N$, $\avee{p^s}$, $\Delta^2 p^s$ and $P^l$ (Tab.~\ref{Tab2}).
Comparing the theoretical and the experimental trajectories of $g^{(2)}(0)$ as a function of $\zeta$, we note an overall good agreement with excellently coinciding values for $\zeta\ge0.6$ within the statistical uncertainties, and slightly deviating trajectories for $\zeta<0.6$. The latter is again attributed to the frequency dependency of the TPA absorption parameter $\beta(\omega)$ of the photomultiplier which does not allow to detect ideal values of $g^{(2)}(0)=2$ in this ultrabroadband emission regime of the QD-SLD, therefore resulting in an experimentally obtained parabola trend $g_{\text{exp}}^{(2)}(0)(\zeta)$ with slightly lower bending, most significantly apparent at low values of $\zeta$. Nevertheless, we observe an overall good reproduction of photon statistical behavior in this mixed light experiment by the analytical quantum theoretical considerations based on the superposition of a well-known coherent light state and the assumed PRAG state. We thus deduce that the broadband light states generated by ASE of the QD-SLD are well described by PRAG states.
\begin{table}
\begin{ruledtabular}
    \begin{tabular}{ c  c  c  c  c }
    $\zeta$ & $N$ & $\frac{\Delta^2 p^s}{\avee{p^s}^2}$ & $g^{(2)}_{\text{th}}(0)$ & $g^{(2)}_{\text{exp}}(0)$\\ \hline
    $0.83\pm0.03$ & 1990 & 0.83 & 1.276 & $1.28\pm0.03$  \\ 
    $0.34\pm0.03$ & 1990 & 0.83 & 1.862 & $1.79\pm0.04$  \\ 
    \end{tabular}
		\end{ruledtabular}
    \caption{Selected, experimentally determined values taken from two measurements of Fig.~\ref{g2exp_theo} and calcultated theoretical counterparts according to our model (Eq. \eqref{g20allgZeta}) for direct comparison to values of the OFB-experiment (Tab.~\ref{Tab1})}
    \label{Tab2}
\end{table}

\section{Conclusion}
\label{conc}
In conclusion, we have studied ultrabroadband amplified spontaneous emission generated by quantum dot superluminescent diodes (QD-SLD) in terms of first and second order correlations, as well as mixing it with coherent light.

For the analysis of the experiments, we considered an $N$-mode phase-randomized Gaussian (PRAG) state. This state is an incoherent superposition of thermal Gaussian states shifted by a complex amplitude $\gamma_i$ for each mode. This ansatz is well suited to match any given near-infrared optical spectrum: it reflects the incoherent character of these broadband emitters and reproduces correct intensity correlations.
We have derived analytical expressions for first and second order correlation functions $g^{(1)}(\tau)$, $g^{(2)}(\tau)$ and $g^{(2)}(0)$, the latter being the footprint of the photon statistics. The intensity correlation $g^{(2)}(0)$ depends functionally on 
the first order correlation $g^{(1)}(\tau)$ with additional finite mode number corrections. 

By a straightforward extension of an optical feedback experiment \cite{Hartmann2013}, we could change the number of modes $N$ by narrowing the spectrum. This resulted in a coherence transition, as seen in Fig.~\ref{feedback}, and agreed very well with the predictions for $g^{(2)}(0)$ by the PRAG state. 

The drawback of spectral narrowing was rectified
by a second experiment creating a mixed light state. 
There, we have superimposed coherent light from a single-mode laser with steady state broadband QD-SLD emission.
As a main result, we obtained broad-range tunable photon statistics, which represents, to the best of our knowledge, the first realization of the mixed light phenomenon including a completely incoherent light component, \ie strong incoherence in both, first and second order correlations (Fig.~\ref{g2exp_theo}).
All relevant experimental features  of the mixed light state can be accounted for with the PRAG state, including the 
temporal correlation functions $g^{(2)}(\tau)$, applicable for pure QD-SLD emission as well as for mixed light at ultra-short timescales (cf. 
Figs.~\ref{tpa-interferogram1} and \ref{tpa_interferomixed}). 

This comprehensive theoretical and experimental study of two different types of tunable photon statistic experiments, validates the simple PRAG state ansatz for broadband QD-SLD ASE. 
This allows us to identify relevant parameters, such as the number of modes $N$ and the statistical properties of their spectral distribution 
$p^s(\omega)$, $\avee{p^s}$ as well as $\Delta^2 p^s$. Future microscopic modeling
of the QD-SLD semiconductor will benefit from these insights.

\section{Acknowledgment}
We like to thank T. Mohr for experimental support and S. V\'{a}rro for fruitful discussions very much. Moreover, we are grateful to R. Phelan (Eblana Photonics) for providing excellent single mode laser devices. We also acknowledge device fabrication and processing from 
I. Krestnikov (Innolume GmbH), M. Krakowski (Thales III-V Lab) and M. Hopkinson (University of Sheffield) within the framework of the EU projects NanoUB-Sources and FastDot.

\appendix

\section{Fitting the optical power spectrum}
\label{fita}
A smooth interpolation of the optical power spectrum $S(\omega)$ of a QD-SLD, which is depicted in Fig.~\ref{spectrum}, is given by a sum of three Gaussian distributions,
\begin{align}
 \label{fit}
 S(\omega)=\sum_{i=1}^3 S_i^0 e^{-\frac{(\omega-\bar\omega_i)^2}{2\sigma_i^2}}.
\end{align}
The numerical data of the fitted amplitudes $S_i^0$, the central frequencies $\bar\omega_i$ and standard deviations $\sigma_i$ are 
tabulated in Tab.~\ref{tablefit}.
\begin{table}[h!]
\begin{ruledtabular}
    \begin{tabular}{c c  c c }
    $i$ & 1 (dashed) & 2 (dashed-dotted) & 3 (dotted)\\ \hline 
    $\bar\omega_i \quad (\unit[]{THz})$ & 2$\pi\cdot242.55$ & 2$\pi\cdot246.05$ & 2$\pi\cdot236.82$ \\ \hline
    $\sigma_i \quad (\unit[]{THz})$ & 2$\pi\cdot2.468$ & 2$\pi\cdot2.875$ & 2$\pi\cdot2.105$ \\ \hline
    $S_i^0\quad (\unit[]{arbitrary~units})$ & 1.904 & 0.637 & 0.532 \\  
    \end{tabular}
\end{ruledtabular}
    \caption{Fit parameters of the Gaussian fit \eqref{fit} to the QD-SLD spectrum depicted in Fig.~\ref{spectrum}.
    \label{tablefit}}
    \end{table}

In this paper, the central frequency of an optical power spectrum $S(\omega)$ is defined as the integral
\begin{align}
\label{meanwidth}
  \bar\omega=\int_{-\infty}^\infty d\omega~\omega~ s(\omega),\quad\text{with}\quad
  s(\omega)=\frac{S(\omega)}{\int_{-\infty}^{\infty}d\omega~ S(\omega)}.
 \end{align}
Consequently, the spectrum in Fig.~\ref{spectrum}, described by the Gaussian distribution $S(\omega)$ in Eq.~\eqref{fit} with the specified parameters of Tab.~\ref{tablefit}, has a central angular frequency $\bar\omega=\unit[2\pi\cdot242.6]{THz}$ or a central wavelength $\bar\lambda=\unit[1236]{nm}$.

A well established definition of the spectral width $\tilde{b}$ is given by the twofold standard deviation,
\begin{align}
 \label{width2}
  \tilde{b}=2\sigma \quad\text{with}\quad\sigma^2=\int_{-\infty}^\infty  d\omega~(\omega-\bar\omega)^2~s(\omega).
\end{align}
The resulting spectral width for the considered spectrum reads 
 $\tilde{b}=\unit[2\pi\cdot7.5]{THz}$.     
 
Generally speaking, for fat-tailed distributions like Lorentzian spectra, the definition of a width in Eq.~\eqref{width2} is not applicable. Therefore, we use an alternative definition for the spectral width
\begin{gather}
    \label{widthmandel}
      b=\frac{1}{\int_{-\infty}^{\infty}d\omega~s^2(\omega)},
\end{gather}
according to \cite{MandelWolf}, also known as Süssmann measure \cite{Schleich}.
In case of a single normalized Gaussian distributed $s(\omega)$ with standard deviation $\sigma$, the spectral width,
\begin{align}
  b_{\text{gauss}}=2\sqrt{\pi}\sigma,
\end{align}
is given by $\sigma$ multiplied by $2\sqrt{\pi}$, \ie a deviation of a factor $\sqrt{\pi}\approx1.77$ compared to the first definition (Eq.~\eqref{width2}). For a spectrum described by Eq.~\eqref{fit} and Tab.~\ref{tablefit}, one obtains $b=\unit[2\pi\cdot13]{THz}$.

Comparing the two definitions of spectral widths, $b=\unit[2\pi\cdot13]{THz}$ (Eq.~\eqref{widthmandel}), $\tilde{b}=\unit[2\pi\cdot7.5]{THz}$ (Eq.~\eqref{width2}) underling different definitions, exhibits a systematic bias. Accordingly, it is important to specify the chosen definition, especially for broadband sources.

\section{Euler-Maclaurin approximation}
The Euler-Maclaurin formula 
approximates a sum by its integral representation and higher order corrections
\begin{gather}
 \label{Euler}
 \sum_{i=1}^N f(a+(i-1)\Delta)=
\frac{1}{\Delta}\int_{a}^b dt f(t)
+\frac{f(a)+f(b)}{2}\\
 +\sum_{m=1}^{M-1}\frac{\Delta^{2m-1}B_{2m}}{(2m)!}\left(f^{(2m-1)}(b)-f^{(2m-1)}(a)\right)+R_M.\nonumber
\end{gather}
Provided that the procedure leads to a vanishing  residual $R_M$, we obtain a series approximation of order $M$ in terms of Bernoulli numbers $B_{k}$ and the higher order derivatives of a function $f^{(k)}$.
The width of the $N-1$ equally spaced integration intervals 
is $\Delta\equiv (b-a)/(N-1)$
\cite{whittaker1927,abramowitzstegun}.

\bibliographystyle{apsrev4-1}
\bibliography{bibliography,textbooks} 
\end{document}